# Semi-analytical Model of Laser Resonance Absorption in Plasmas

**S J Pestehe\* and M Mohammadnejad**

Department of Physics, University of Tabriz, Tabriz, Iran

\* Fax: +98(411)3341244, Email: sjpest@tabrizu.ac.ir

**Abstract**. When an electromagnetic wave is obliquely incident on an inhomogeneous high density plasma, it will be absorbed resonantly as long as it is polarized in the plane of incidence and has an electric field component along the plasma electron density gradient. This process takes place by linear mode conversion into an electron plasma wave. In this paper, we have considered the resonant absorption of laser light near the critical density of a plasma with linear electron density profile. The behaviour of the electric and magnetic vectors of a laser light propagating through inhomogeneous plasma has been studied by calculating them using Maxwell's equations using a new semi-analytical model. The absorbed fraction of the laser light energy, then, evaluated and plotted versus the angle of incidence. It has been shown that this new model can explain the previous classical approximated results at high density scale lengths as well as the reported numerical results in almost all density scale lengths.

## 1. Introduction

It is well known that an electromagnetic wave obliquely propagating through inhomogeneous high density plasmas with its electric field placed on the plane of incidence will find a singularity at a point where the permittivity of the plasma vanishes and so will be absorbed resonantly. This process takes place by resonantly driving large amplitude plasma electron wave. Damping of this resonantly derived plasma wave occurs both by collisional and non-collisional processes. There are several works that have considered different aspects of this phenomenon. Some of the early workers studied the propagation of the electromagnetic waves through the cold plasmas neglecting any plasma wave damping effects [1]. Denisove [2] and Piliya [3] have extended this earlier works. The absorption coefficient for warm plasmas in the low temperature regime was the main result of Piliya's mathematical analysis which has been re-extended by Kelly and Banos [4] and Forslund et al [5] using numerical techniques to calculate the wave solution. Their results which, were in disagreement with the results of Piliya has been re-examined by Pert [6] to resolve this discrepancy. Hammerling [7] presented an angle averaged resonant absorption function to take into account of the experimental conditions where the angles of incidence vary greatly and in unknown manner. Almost in all of these works and others the electron plasma density profiles are assumed to be linear and the absorption fraction for the resonance absorption is approximated near the critical surface.

In this paper we have also considered resonant absorption of laser light near the critical density of plasmas with linear electron density profile. But a quite different and new method is used to calculate the linear resonant absorption of the laser light near the critical surface. The electric and magnetic field vectors of the laser light propagating through the plasma are calculated by solving Maxwell's equations in inhomogeneous plasma with the assumption that the space can be divided to three sharply distinct regions such as, the vacuum (outside the plasma), inside the plasma from a vacuum-plasma interface up to the critical surface with a linear electron density, and a high density region beyond the critical surface with a constant electron density. Absorbed fraction of the energy of the laser light is also evaluated and plotted versus the angle of incidence and related parameters. It has been shown that our new semi analytical model is able to explain the previous classical approximated results at high density scale length as well as the reported numerical results of Hong-bo Cai [8] in almost all density scale length.

2. **The model and field calculations**

We consider a plane parallel inhomogeneous plasma whose density gradient is in the x direction. It is assumed that an electromagnetic wave of frequency $\omega_0$ and initial wave vector $k_0$ having its wave vector on (x-y) plane, incident on the vacuum-plasma interface at an angle of $\theta_0$ with respect to the x axis. The electric field of the wave, therefore, will have only z component for the S polarization and both of x and y components for the P polarization cases and vice versa for the magnetic field of the electromagnetic wave.

Considering an S polarized electromagnetic wave obliquely incident on the plasma and propagating along the x direction, the electric field of the wave can be given by

$$E_z(x,y,t) = E(x)\exp[i(k_0 \sin\theta_0 y - \omega_0 t)] \tag{1}$$

where $k_0 = \dfrac{\omega_0}{c}$ is the wave number of the electromagnetic field in free space. The wave equation for that electric field is

$$\frac{d^2 E(x)}{dx^2} + k_0^2 \left[\varepsilon(x) - \sin\theta_0^2\right]E(x) = 0 \quad. \tag{2}$$

To solve this equation we assume that the space is consisting of three sharply distinct regions such as (a) vacuum with $\varepsilon = 1$, (b) inside the plasma from vacuum-plasma interface to the critical surface with a linear electron density profile such as

$$n_e = n_c \frac{x}{L}$$

with

$$\varepsilon = 1 - \frac{n_e}{n_c} = 1 - \frac{x}{L} \tag{3}$$

and (c) the high density region beyond the critical surface. In this case it is assumed that the electron density remains constant after the $x = NL$, then

$$\varepsilon = 1 - \frac{n_e}{n_c} = 1 - N \tag{4}$$

where $n_c = \dfrac{m\omega_0^2}{4\pi e^2}$ is the critical density of the plasma and $N \geq 1$ is a positive number which will be taken to be 1.5 throughout this work. The assumed plasma electron density profile is summarized in the following equation:

$$\begin{cases} n_e = 0 & \text{for } x \leq 0 \\ n_e = n_c \dfrac{x}{L} & \text{for } 0 < x < NL \\ n_e = n_c N & \text{for } x \geq NL \end{cases}$$

Substituting permittivity relations into equation (2) the wave equations for these three regions, respectively, are

$$\frac{d^2 E(x)}{dx^2} + k_0^2 \cos\theta_0^2 E(x) = 0 \qquad (5)$$

$$\frac{d^2 E(x)}{dx^2} + k_0^2 \left[1 - \frac{x}{L} - \sin\theta_0^2\right] E(x) = 0 \qquad (6)$$

$$\frac{d^2 E(x)}{dx^2} - k_0^2 \left[N - \cos\theta_0^2\right] E(x) = 0 \qquad (7)$$

Solutions that fulfil the boundary conditions, respectively, are

$$E_1(x) = E_0 \sin(k_0 \cos\theta_0 x + \varphi) \qquad (8)$$

$$E_2(\eta) = \alpha A_i(\eta) + \beta B_i(\eta) \qquad (9)$$

$$E_3(x) = \gamma \exp(-\kappa k_0 x) \qquad (10)$$

where $A_i(\eta)$ and $B_i(\eta)$ are the Airy functions and $\varphi$ is the phase of the wave when incident on the vacuum-plasma interface and $\eta$ and $\kappa$ are defined as

$$\eta = \left(\frac{k_0^2}{L}\right)^{\frac{1}{3}} \left(x - L\cos^2\theta_0\right), \quad \kappa = \sqrt{N - \cos\theta_0^2} \qquad (11)$$

where $\alpha, \beta$, and $\gamma$ are constants which should be determined using boundary conditions.

Using Maxwell's equations we can relate the magnetic field to the electric field as

$$\vec{B}(x, y) = \frac{i}{k_0}\left(ik_0 \sin\theta_0 \vec{E}(x)\hat{x} - \frac{d\vec{E}}{dx}\hat{y}\right)\exp(ik_0 \sin\theta_0 y) \qquad (12)$$

Therefore the magnetic fields at the three defined regions, respectively, given by

$$\begin{aligned}\vec{B}_1(x, y) = -E_0[&\sin\theta_0 \sin(k_0 \cos\theta_0 x + \varphi)\cos(k_0 \cos\theta_0 y)\hat{x} \\ -&\cos\theta_0 \cos(k_0 \cos\theta_0 x + \varphi)\sin(k_0 \cos\theta_0 y)\hat{y}]\end{aligned} \qquad (13)$$

$$\vec{B}_2(x,y) = -\{\sin\theta_0 \left[\alpha A'_i(\eta) + \beta B'_i(\eta)\right]\cos(k_0 \sin\theta_0 y)\hat{x} \\ - (k_0 L)^{-\frac{1}{3}}\left[\alpha A'_i(\eta) + \beta B'_i(\eta)\right]\sin(k_0 \sin\theta_0 y)\hat{y}\} \qquad (14)$$

$$\vec{B}_3(x,y) = -\gamma\left(\sin\theta_0 \cos(k_0 \sin\theta_0 y)\hat{x} - \kappa \sin(k_0 \sin\theta_0 y)\hat{y}\right)\exp(-\kappa k_0 x) \qquad (15)$$

Applying the boundary conditions at $x = 0$ and $x = NL$ namely, the continuity of the tangential components of the electric and magnetic fields at boundaries (the later fulfils as there is no currents at boundaries) and defining two new parameters such as $\eta_0 = -(k_0 L)^{\frac{2}{3}}\cos^2\theta_0$ and $\eta_1 = (k_0 L)^{\frac{2}{3}}\kappa^2$ one gets

$$E_0 \sin\varphi = \alpha A_i(\eta_0) + \beta B_i(\eta_0) \qquad (16)$$

$$\alpha A_i(\eta_1) + \beta B_i(\eta_1) = \gamma \exp[-\kappa k_0 (NL)] \qquad (17)$$

$$E_0 \cos\theta_0 (k_0 L)^{\frac{1}{3}}\cos\varphi = \alpha A'_i(\eta_0) + \beta B'_i(\eta_0) \qquad (18)$$

$$\alpha A'_i(\eta_1) + \beta B'_i(\eta_1) = -\gamma\kappa(k_0 L)^{\frac{1}{3}}\exp[-\kappa k_0 (NL)] \qquad (19)$$

which can be solved easily to obtain

$$\beta = C_1 \alpha,\quad C_1 = -\frac{\kappa(k_0 L)^{\frac{1}{3}} A_i(\eta_1) + A'_i(\eta_1)}{\kappa(k_0 L)^{\frac{1}{3}} B_i(\eta_1) + B'_i(\eta_1)} \qquad (20)$$

$$\alpha = E_0 \left(\left[A_i(\eta_0) + C_1 B_i(\eta_0)\right]^2 - \frac{\left[A'_i(\eta_0) + C_1 B'_i(\eta_0)\right]^2}{\eta_0}\right)^{-\frac{1}{2}} \qquad (21)$$

$$\varphi = \sin^{-1}\left[\frac{\alpha A_i(\eta_0) + \beta B_i(\eta_0)}{E_0}\right] \qquad (22)$$

$$\gamma = e^{\kappa(k_0 L)N}\alpha A_i(\eta_1) + \beta B_i(\eta_1) \qquad (23)$$

Substituting the corresponding relations for $\alpha$, $\beta$, $\gamma$, and $\varphi$ in equations (13), (14), and (15) the magnetic fields at the defined three regions are determined.

The behaviour of the electric and magnetic fields with distance along their propagation directions for a typical values of $k_0 L = 10$ and $\theta_0 = 50°$ are shown in figures (1) and (2), respectively.

For the purpose of energy absorption calculations we will need to have the magnetic field vector and its magnitude at the reflection point $x = L\cos^2\theta_0$, where $\eta = 0$. We obtain

$$\vec{B}_2\left(x = L\cos^2\theta_0\right) = \sin\theta_0\left[\alpha A_i(0) + \beta B_i(0)\right]\hat{x} \\ -\left(k_0 L\right)^{-\frac{1}{3}}\left[\alpha A_i'(0) + \beta B_i'(0)\right]\hat{y} \quad (24)$$

for the magnetic field vector at the reflection point and

$$\left|B_2\left(x = L\cos^2\theta_0\right)\right| = \sqrt{\Sigma} \quad (25)$$

where

$$\Sigma = \left(k_0 L\right)^{\frac{2}{3}}\left[\alpha A_i'(0) + \beta B_i'(0)\right]^2 + \sin^2\theta_0\left[\alpha A_i(0) + \beta B_i(0)\right]^2$$

for the magnitude of the magnetic field at that point.

### 3. Fractional absorption

Let us concentrate on the second region from the vacuum-plasma interface until the critical surface. The relation between the magnitude of the x component of the electric field and magnitude of the magnetic field is

$$E_x = \frac{\sin\theta_0 B(x)}{\varepsilon(x)} \quad . \quad (26)$$

Since $E_x$ has a singularity at the critical point therefore, following Kruer [9] we define the resonantly driven field $E_l = \sin\theta_0 B(x)$ which can be interpreted as a component of the electric field of the laser light causing electrons oscillate along the electron density gradient. It has to be mentioned that this resonantly driven field is evaluated at the resonant point.

Assuming a small damping with a frequency of $\nu$ to the wave, the dielectric function can be approximated as:

$$\varepsilon(x) = 1 - \frac{\omega_p^2}{\omega_0^2}(1 + i\frac{\nu}{\omega_0})^{-1} \quad . \tag{27}$$

The absorbed fraction of the laser energy, $f_a$, is given by [9]:

$$f_a = \frac{\nu}{8\pi I_0} \int_{-\infty}^{+\infty} \frac{E_l^2(x)}{|\varepsilon(x)|^2} dx \tag{28}$$

Assuming that the resonant region is narrow, the electric field can not change appreciably across it and so, we can take it out if the integration and write the absorption fraction as:

$$f_a = \frac{\nu}{8\pi I_0} E_l^2 \int_{-\infty}^{+\infty} \frac{1}{|\varepsilon(x)|^2} dx \tag{29}$$

We take the magnitude of the magnetic field at the reflection point for the S polarized obliquely incident light the same as it's magnitude for the P polarized one at that point and relate it to the magnetic field at the critical density position as [9]

$$B_2(x = L) \approx B_2(x = L\cos^2\theta_0)\exp(-\delta) \tag{30}$$

where $\delta$ is the magnetic field amplitude absorption coefficient and is given by

$$\delta = \int_{L\cos^2\theta_0}^{L} \frac{1}{c}\sqrt{\omega_{pe}^2 - \omega^2\cos^2\theta_0}\, dx = \frac{2k_0 L\sin^3\theta_0}{3} \tag{31}$$

Substituting $\delta$ and magnitude of $B_2(x = L\cos^2\theta_0)$ from equations (31) and (25), respectively, into equation (30), the magnetic field at the critical density is obtained

$$B_2(x = L) = \sqrt{\Sigma}\exp(-\frac{2}{3}k_0 L\sin^3\theta_0) \tag{32}$$

The electric field at the resonant point, then, is given by:

$$E_l(x=L) = (k_0 L)^{-\frac{1}{3}} \tau \exp(-\frac{2}{3}\tau^3)\sqrt{\Sigma} \qquad (33)$$

where the parameter $\tau = (k_0 L)^{\frac{1}{3}} \sin\theta_0$ has been used.

Using equations (29) and (33) together with the linear electron density profile, the absorption fraction becomes

$$f_a = \frac{\nu E_l^2(x=L)}{c E_0^2} \int_{-\infty}^{+\infty} \left( (1-\frac{x}{L})^2 + \frac{\nu^2}{\omega^2}\frac{x^2}{L^2} \right)^{-1} dx \qquad (34)$$

Calculating the integration with the assumption of $\frac{\nu}{\omega_0} \ll 1$ we get

$$f_a = \frac{\pi (k_0 L) E_l^2(x=L)}{E_0^2} \qquad (35)$$

The fractional absorption for typical values of $k_0 L$ is plotted versus $\tau = (k_0 L)^{\frac{1}{3}} \sin\theta_0$ in figure (3). We have also shown the numerical results reported by Cai [8] in figure (4). It can be seen that they both behave in a very good agreement but the peaks in figure (5) are a little right shifted along the $\tau$ axis.

The fractional absorptions obtained from our model together with Kruer's [9] (reproduced here) are plotted versus incident angle for three typical different values of $k_0 L = 10$, 1, and 0.5 in figures (5), (6) and (7), respectively. It can be seen that the optimum incident angle for the two models coincide for large $k_0 L$s namely $k_0 L = 10$. But it is evident that for small scale length like $k_0 L = 0.5$, Kruer's model give the optimum incident angle of nearly 90° which is physically not acceptable while our new model lead to 50°.

The optimum incident angle and the fractional absorption are illustrated as functions of $k_0 L$ in figure (8). It can be seen that the laser absorption increases with the increasing density scale length , $k_0 L$, while the optimum angle decreases with it.

We have also compared the optimum angle obtained from our model with Kruer's, by plotting them versus $k_0 L$ in figure (9) which shows their disagreement at the small density scale lengths.

## 4. Conclusions

The interaction of laser light with plasmas having linear electron density profiles via the resonance absorption process is studied. The behaviours of the electric and magnetic fields of the laser light propagating through the plasma have been studied by dividing the space into three different sharply distinct regions and solving the corresponding wave equations in those regions and then relating the obtained solutions using boundary conditions.

The fractional absorption has been calculated and showed that the new analytical model can explain the classical approximated Kruer's result at high density scale length as well as the numerical results of Cai in almost all density scale length for the linear density profile. It has been shown that the maximum energy transfer to the plasma, with linear density profile, occurs at higher $\tau = (k_0 L)^{\frac{1}{3}} \sin \theta_0$ with increasing $k_0 L$. It has been shown that nearly 45% laser energy transfer to the plasma at the critical surface is obtainable at $k_0 L = 10$ with $\tau = 0.9$. The study of the dependencies of the optimum laser incident angle and the fractional absorption, to the plasma scale lengths, $k_0 L$, has shown that the laser absorption increases with the increasing density scale length while the optimum angle decreases. Comparing the optimum incident angle obtained from our new model with the corresponding results from Kruer, by investigating their dependencies to the plasma scale lengths has shown their disagreement at the small density scale lengths.

**Acknowledgements**

General financial support of the Urmia University of Iran for this work is gratefully acknowledged.

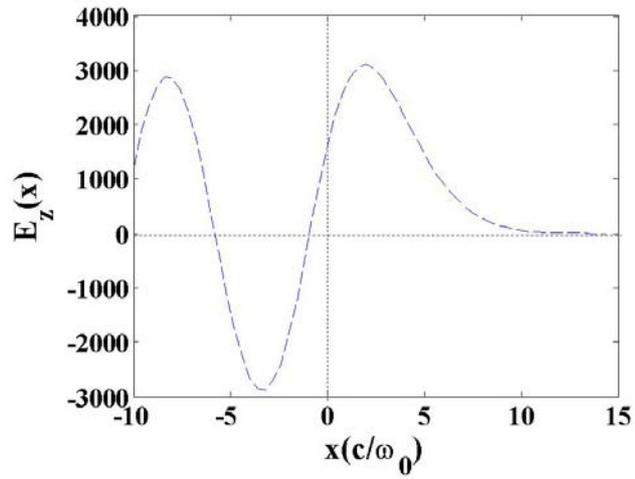

Figure 1) The behaviour of the electric field, propagating through a plasma with linear density profile, with distance for typical values of $k_0 L = 10$ and $\theta_0 = 50°$.

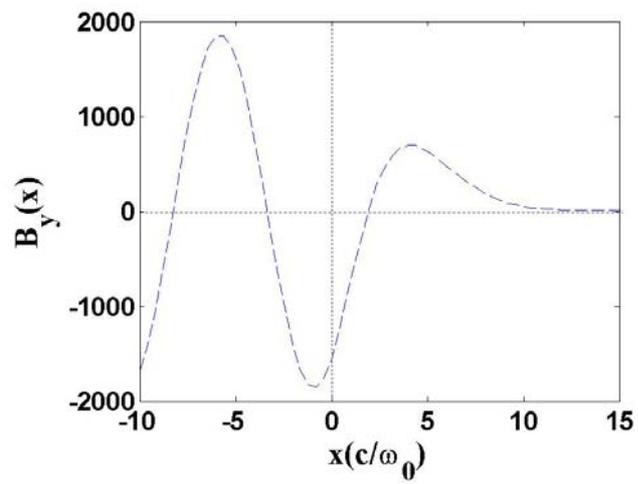

Figure 2) The behaviour of the magnetic field propagating through a linear density plasma with distance for some typical values of $k_0 L = 10$ and $\theta_0 = 50°$

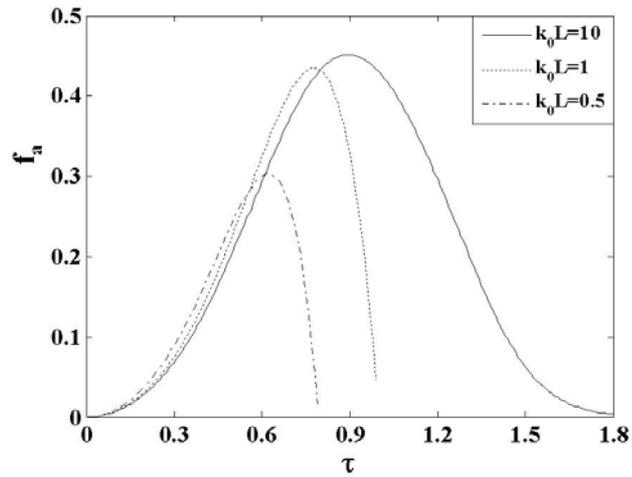

Figure 3) The fractional absorption for different $k_0 L$ s obtained using the new semi analytical model. It can be seen that the peak absorption occurs at higher $\tau$ s, as $k_0 L$ increases.

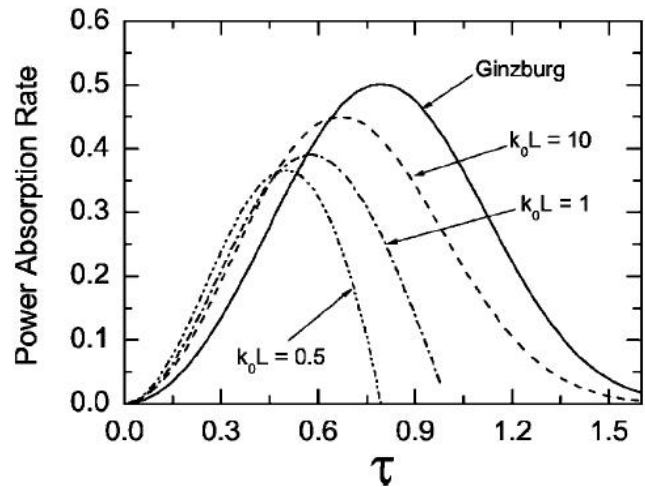

Figure 4) The absorption fraction for different $k_0 L$ s reported as the numerical results by Hong-bo Cai[8]

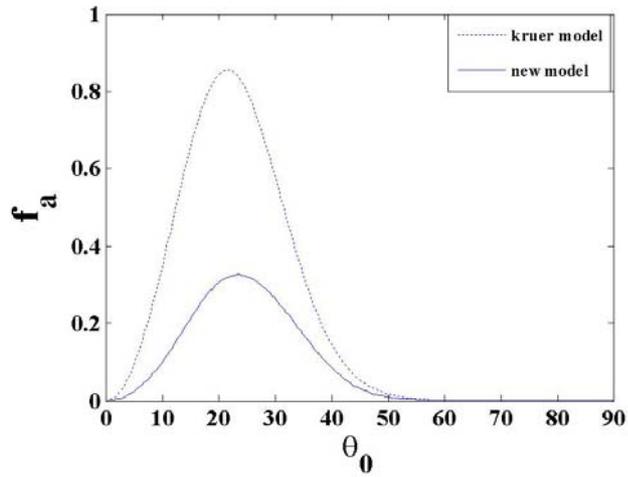

Figure 5) Comparing the results for absorption fraction obtained by Kruer [9] with the results from our new model for $k_0 L = 10$

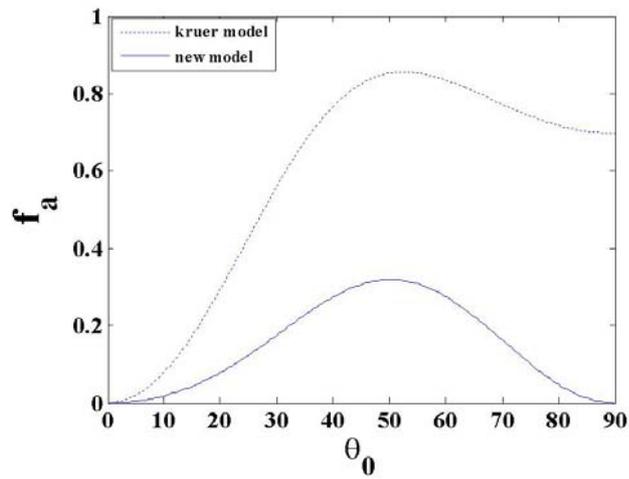

Figure 6) Absorption fraction obtained with the new model for $k_0 L = 1$ with corresponding results from Kruer [9], (reproduced here)

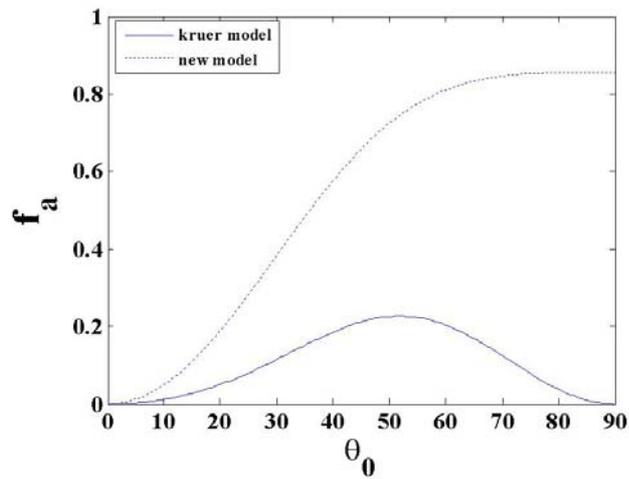

Figure 7) The Comparison of the absorption fraction obtained from the Kruer's [9] model with the results from our model for $k_0 L = 0.5$.

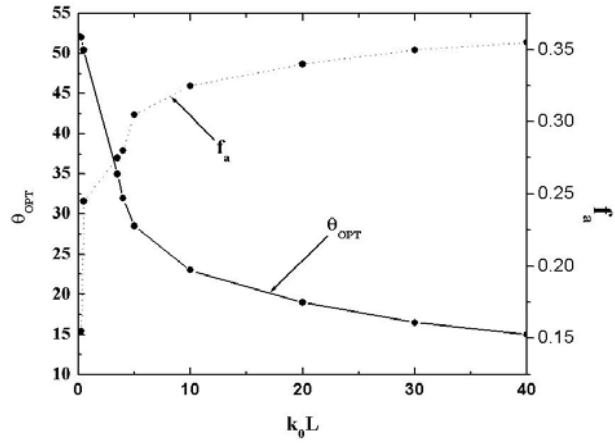

Figure 8) Variations of the optimum angle and the fractional absorption versus $k_0 L$. The laser absorption increases with the increasing density scale length ($k_0 L$) while the optimum angle decreases.

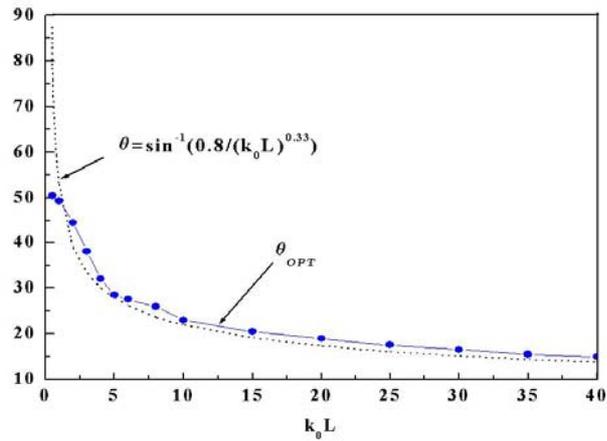

Figure 9) Optimum angle as a function of $k_0 L$. The dashed line is for Kruer's classical results and the solid circle is for optimum angle from the new model, respectively.